\renewcommand\a{\alpha}

\newcommand{\diracslash}[1]{#1\llap{/\kern2pt}}

\newcommand{\be}{\begin{equation}}
\newcommand{\ee}{\end{equation}}
\newcommand{\bea}{\begin{eqnarray}}
\newcommand{\eea}{\end{eqnarray}}
\newcommand{\ba}[1]{\begin{array}{#1}}
\newcommand{\ea}{\end{array}}

\documentclass[twocolumn,superscriptaddress,secnumarabic,amssymb,nobibnotes,nofootinbib,aps,pre,showpacs]{revtex4}
\usepackage{graphicx}
\begin{document}
\setlength{\topmargin}{-0.05in}

\title{Parametrically controlling solitary wave dynamics in modified Kortweg-de Vries equation
}

\author{Kallol Pradhan}
\affiliation{Physics Department, University of Wisconsin-Madison,
1150 University Avenue, Madison, WI53706 -1390, U.S.A.}

\author{Prasanta K. Panigrahi}\thanks{prasanta@prl.res.in}\affiliation{Physical
Research Laboratory, Navrangpura, Ahmedabad 380 009, India}

\date{\today}

\begin{abstract}
We demonstrate the control of solitary wave dynamics of modified
Kortweg-de Vries (MKdV) equation through the temporal variations
of the distributed coefficients. This is explicated through exact
cnoidal wave and localized soliton solutions of the MKdV equation
with variable coefficients. The solitons can be accelerated and
their propagation can be manipulated by suitable variations of the
above parameters. In sharp contrast with nonlinear Schr\"{o}dinger
equation, the soliton amplitude and widths are time independent.
\end{abstract}

 \maketitle

Modified Kortweg-de Vries (MKdV) equation manifests in diverse areas
of physics \cite{Miura,Wadati1,Wadati2,Hirota,Zuntao,Matsutani}. For
example, it appears in the context of, electromagnetic waves in
size-quantized films, van Alfv\'{e}n waves in collisionless plasma
\cite{khater}, phonons in anharmonic lattice \cite{ono}, interfacial
waves in two layer liquid with gradually varying depth
\cite{helfrich}, transmission lines in Schottky barrier
\cite{ziegler}, ion acoustic solitons
\cite{lonngren,watanabe,tajiri}, elastic media \cite{Tsuru}, and
traffic flow problems \cite{takashi,komatsu}. It is an integrable
dynamical system with an infinite number of conserved quantities;
the solutions of this equation are well studied \cite{book1,Khare}.

Recently non-linear equations with variable coefficients have
attracted considerable attention in the literature. Nonlinear
Schr\"{o}dinger equation (NLSE) with variable non-linearity and
dispersion is relevant to both optical fibers and Bose-Einstein
condensates \cite{moors,kruglov,hasegawa,Atre1}. Nonlinear
Schr\"{o}dinger equation with source, having distributed
coefficients like variable dispersion, variable Kerr nonlinearity
and gain or loss, is applicable to asymmetric twin-core optical
fibers \cite{raju1,raju2}. It has been shown that, solitons can be
compressed and their dynamics effectively controlled through these
variable parameters. The Kortweg-de Vries (KdV) equation with
variable coefficients \cite{tian} has been studied recently in the
context of ocean waves, where the spatio-temporal variability of
the coefficients are due to the changes in the water depth and
other physical conditions. The fact that, MKdV equation is
relevant to hydrodynamics and a variety of physical phenomena, it
is natural to expect the possibility of temporal variations in the
equation parameters occurring in the same. Furthermore, for
propagating solitons, the first integral of the MKdV equation
yields NLSE with a source, making it imperative to investigate the
effect of the temporal variation of the distributed parameters on
the solitary wave solutions of this dynamical system.

The goal of the present paper is to study the effect of the
variable coefficients on the solution space of MKdV equation, both
for positive and negative cases. We find that the effect of
distributed coefficients on the soliton dynamics of MKdV equation
is quite different than that of NLSE. In case of NLSE, the
amplitude and width are affected by the time dependence of the
distributed coefficients. This leads to compression of solitary
waves in NLSE. Through explicit construction, it is shown that,
solitary waves in this system can be effectively controlled
through the equation parameters. The solitons can be accelerated
and manipulated by suitable variations of the above parameters.
However the width and amplitudes are not amenable for manipulation
and control.

We consider the modified KdV equation with variable coefficients
in the form
 \begin{equation} \label{mkdv}
u_{t}+\alpha(t)u_x-\beta(t)u^2u_x+\gamma(t)u_{xxx}=0,
\end{equation}where $\gamma(t)$, $\alpha(t)$ and
$\beta(t)$ are time dependent variables. Although, the first
derivative term in the field variable can be removed by a suitable
change of the coordinate frame, the same has been kept here
explicitly to contrast its effect with the nonlinear and
dispersion terms. We consider the ansatz solution of the form,
\begin{equation} \label{ansatz}
u= A_{1}(t)g[{\omega(x,t)}]+A_{0}(t),
\end{equation}
where, ${\omega}(x,t)=f(t)x-h(t)$. The variable coefficient MKdV
equation can be mapped to Jacobi elliptic equation:
\begin{equation} \label{relation1}
g^{''}=Pg + 2Qg^{3},
\end{equation}
having a conserved quantity,
\begin{equation} \label{relation2}
(g^{'})^{2}=Pg^{2} + Qg^{4}.
\end{equation}

{\noindent}Here prime indicates differentiation with respect to
the argument $\omega$, and $P$ and $Q$ are constants. Substituting
the ansatz in Eq.(\ref{mkdv}) and using the relations
(\ref{relation1}) and (\ref{relation2}), we collect the
coefficients of ${g^{\a}}$ and ${g^{\a}g^{'}}$ (where $\a$=0,1,2)
to find the consistency conditions:
\begin{equation} \label{cc1}
g^{0} :  \partial_{t} A_{0}=0,
\end{equation}

\begin{equation} \label{cc2}
g :  \partial_{t} A_{1}=0,
\end{equation}
\begin{equation} \label{cc3}
{\nonumber} g ^{'}:
A_{1}\partial_{t}\omega+A_{1}\alpha(t)\partial_{x}\omega
\\
+A_{1}P\gamma(t)[\partial_{x}\omega]^3-A_{0}^2A_{1}\beta(t)\partial_{x}\omega=0,
\end{equation}
\begin{equation} \label{cc4}
gg ^{'}: -2A_{0}A_{1}^2\beta(t)\partial_{x}\omega=0,
\end{equation}

\begin{equation} \label{cc5}
g^2g^{'}:-A_{1}^3\beta(t)\partial_{x}\omega+6A_{1}Q\gamma(t)[\partial_{x}\omega]^3=0,
\end{equation}
For obtaining solutions to the above set of equations, we require,
$ A_{0}=0$, and $A_{1}=$constant. From Eq.(\ref{cc5}), further
simplification yields,
\begin{equation} \label{relation3}
{f^2}(t)= \frac{ A_{1}^2\beta(t)} {6Q\gamma(t)},
\end{equation} \label{relation4}
and from Eq.(\ref{cc3}) we get,
\begin{equation} \label{relation5}
\partial_{t}h(t)=x \partial_{t}f(t) +\alpha(t)f(t)+P\gamma(t){f^3}(t).
\end{equation}
Condition (\ref{relation5}) requires that, $f(t)$ should be a
constant so that the term containing $x$ vanishes. This implies
$\beta(t)/\gamma(t)=\kappa$, where $\kappa$ is a constant. We then
have the relations,
\begin{equation}
f=\sqrt{ \frac {A_{1}^2 \kappa} {6Q} },
\end{equation}
and,
\begin{equation}
h(t)= \sqrt{ \frac {A_{1}^2 \kappa} {6Q} } \int [ \alpha(t)+ \frac
{\gamma(t)PA_{1}^2\kappa } {6Q} ] dt.
\end{equation}

 Hence, the exact travelling wave  solution can be written in the
form,
\begin{equation} \label{soln}
u=A_{1}g\left( \sqrt {\frac{A_{1}^2\kappa} {6Q}} \left[x-\int
[\alpha(t)+ \frac {\gamma(t)PA_{1}^2\kappa} {6Q} ]dt
\right]\right).
\end{equation}
It is worth noting that, $A_{1}$ is unconstrained and controls the
width of the solution. Unlike the case of NLSE the amplitude and
width are independent of time. Since, the solution involves
$\kappa$, positive and negative MKdV equations have different type
of solutions. g can be any of the twelve Jacobi elliptic
functions, with the modulus parameter $m^2$ $(0\leq m^2 \leq 1)$
\cite{jose, stegun}.
 The following are some
identities of the Jacobi elliptic functions, which are used:
$cn^2(w,m)+sn^2(w,m)=1$,\\ $dn^2(w,m)+m^2sn^2(w,m)=1$,\\
$sn^{'}(w,m)=cn(w,m){\cdot}dn(w,m)$,\\$cn^{'}(w,m)=-sn(w,m){\cdot}dn(w,m)$,\\
$dn^{'}(w,m)=-m^2sn(w,m){\cdot}cn(w,m)$.
\\For
$m=1$,
\\$cn(w,1)=dn(w,1)=sech(w)$ and $sn(w,1)=tanh(w)$.

Below we analyze some explicit solutions and corresponding
parameter ranges. For the sake of specificity we consider
$\beta(t)>0$.

\noindent {\bf{Case I}}: With $g= cn(\omega(x,t))$ one finds, the
cnoidal wave solution as,

\begin{equation}
u=A_{1} cn\left( \sqrt {\frac{A_{1}^2\kappa} {-6m^2}} \left[x-\int
[\alpha(t)+ \frac {\gamma(t)(2 m^2 -1) A_{1}^2\kappa} {-6m^2} ]dt
\right]\right),
\end{equation}

where, $\gamma(t)<0$, $P=(2m^2-1)>0$, $Q=-m^2<0$ and $m^2 > 1/2$.
This corresponds to the positive MKdV equation. In the case, when
$m^2=1/2$, $P=0$, and $\gamma(t)$ does not affect the solution.
 For, $m^2=1$ case, we have the exact
solitary wave solution of the form,
\begin{equation}
u=A_{1} sech\left( \sqrt {\frac{-A_{1}^2\kappa} {6}} \left[x-\int
[\alpha(t)- \frac {\gamma(t) A_{1}^2\kappa} {6} ]dt
\right]\right).
\end{equation}

\begin{figure*}
\begin{tabular}{cc}
\scalebox{0.65}{\includegraphics{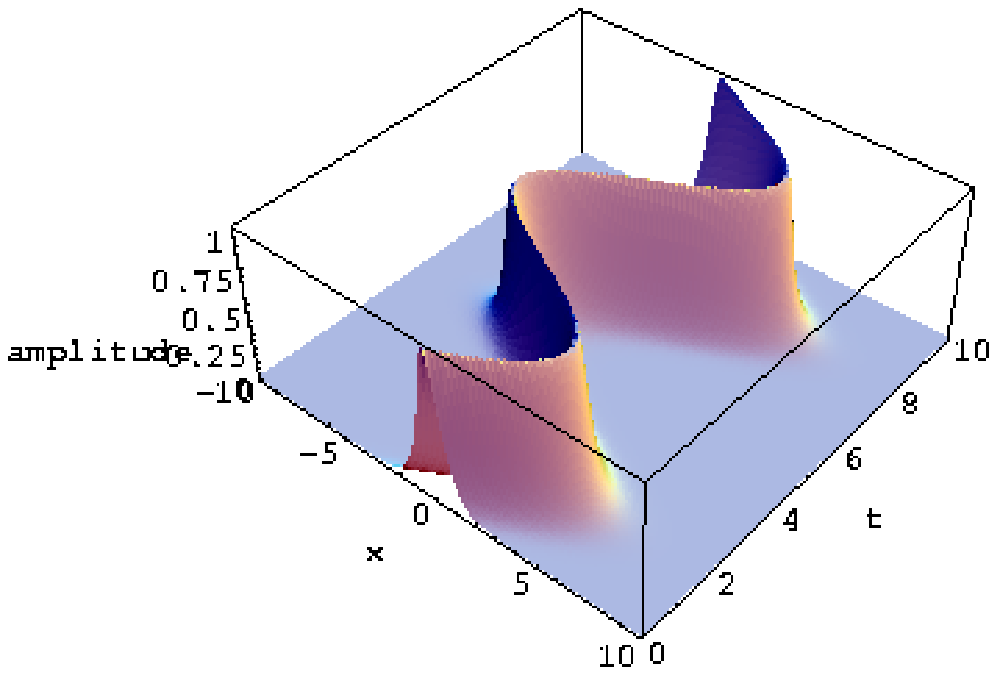}}&
\scalebox{0.65}{\includegraphics{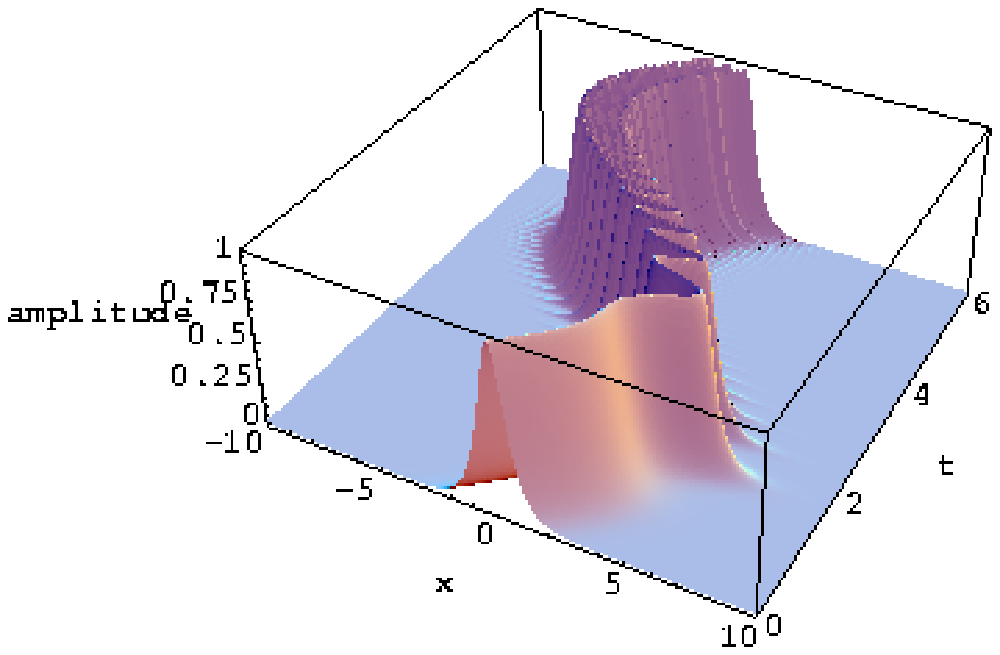}}
\end{tabular}
\caption{\label{fig1} Propagating localized solitary wave solution
of MKdV equation with $g=cn(x,t)$, where $\gamma= \cos(t)$, $
\kappa=-24$, $A_{1}=1$, $P=1$, $Q=-1$ and $m=1$; left $\alpha(t)=0$
and right $\alpha(t)=-3t{^3}\cos(t^3)$.}
\end{figure*}

\begin{figure*}
\begin{tabular}{cc}
\scalebox{0.65}{\includegraphics{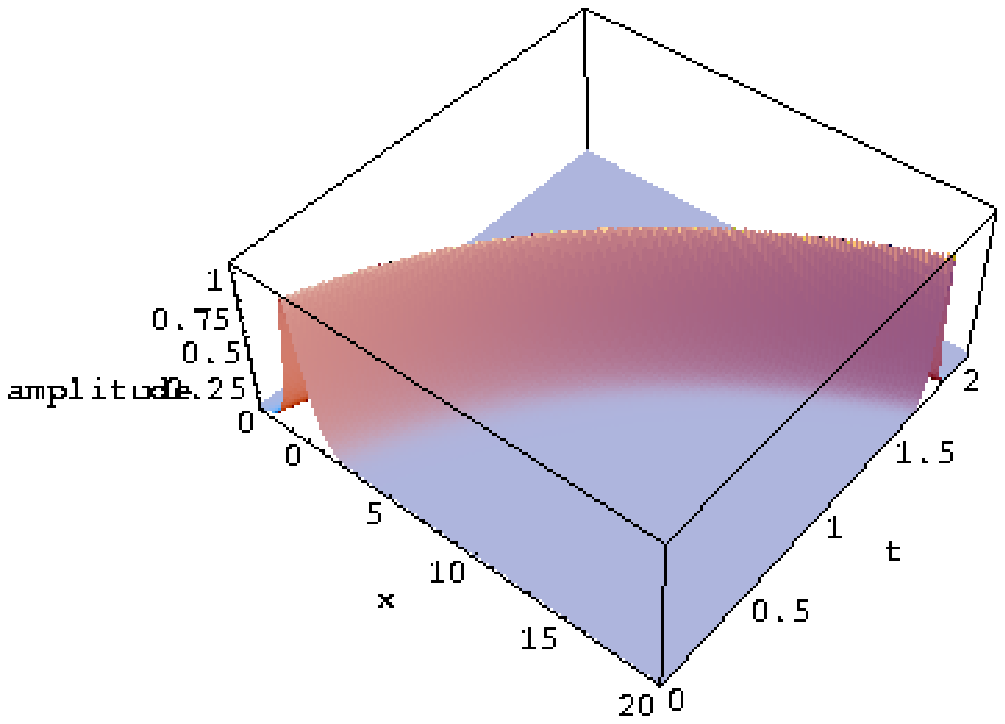}}&
\scalebox{0.65}{\includegraphics{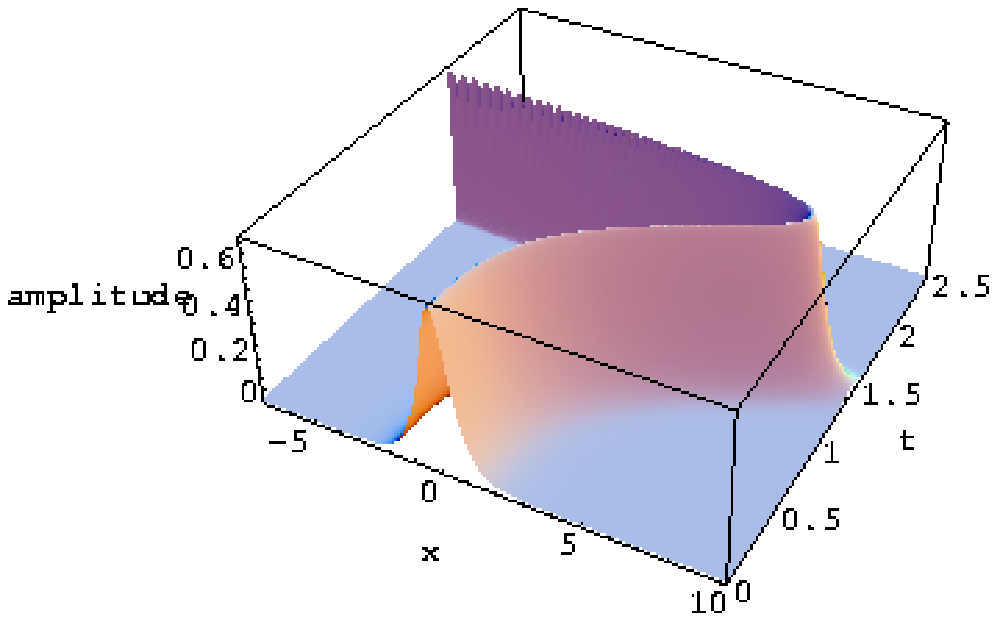}}
\end{tabular}
\caption{\label{fig2} Propagating localized solitary wave solutions
for $g=cn(x,t)$, where $\gamma(t)=3t^2$, $ \kappa=-24$,$A_{1}=1$,
$P=1$, $Q=-1$ and $m=1$; left $\alpha(t)=0$ and right
$\alpha(t)=-t^9$.}
\end{figure*}

Fig. (\ref{fig1}) depicts the temporal evolution of the  above
bell shaped localized solution. For illustrative purpose we have
considered two different cases where, $\alpha(t)=0$ and
$\alpha(t)=-3 t^3 cos(t^3)$. Fig. (\ref{fig2}) depicts the same
solution when $\gamma(t)$ and $\alpha(t)$ have polynomial time
dependence. One clearly sees that the temporal variations of
$\gamma$ and $\alpha$ can effectively modulate and control the
propagation of the solitons.

\noindent {\bf{Case II}}: $g=sn(w,m)$\\

We now study the cases where $g=sn(\omega(x,t))$, for which the
solution corresponds to negative MKdV equation:

\begin{equation}
u=A_{1} sn\left( \sqrt {\frac{A_{1}^2\kappa} {6m^2}} \left[x-\int
[\alpha(t)- \frac { \gamma(t)(m^2 +1) A_{1}^2\kappa} {6m^2} ]dt
\right]\right).
\end{equation}

Here $\gamma(t)>0$, $Q=m^2>0$ and $P=-(m^2 +1)<0$.\\For $m^2=1$,
we have the kink type solitary wave solution
\begin{equation}
u=A_{1} tanh\left( \sqrt {\frac{A_{1}^2\kappa} {6}} \left[x-\int
[\alpha(t)- \frac { \gamma(t) A_{1}^2\kappa} {3} ]dt
\right]\right).
\end{equation}

\begin{figure}
\centering
\includegraphics[width=2.25in]{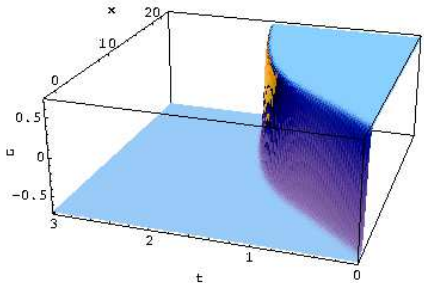}
\caption{ \label{fig3} Kink type solitary wave solution of the MKdV
equation, where $\gamma(t)=\cos(t)$, $\alpha(t)=5t^4$, $\kappa=-48$,
$A_{1}=1$, $P=1$, $Q=-1$ and $m=1$.}
\end{figure}

Fig. (\ref{fig3}) depicts the kink solution in the presence of
time dependent dispersion. The soliton motion can be controlled
through the external parameters.

 \noindent {\bf{Case III}}: $g=dn(w,m)$

We get exact solitary wave solution  of positive MKdV equation, in
the case when $g= dn(w(x,t))$:

\begin{equation}
u=A_{1} dn\left( \sqrt {\frac{-A_{1}^2\kappa} {6}} \left[x-\int
[\alpha(t)- \frac {\gamma(t)(2-m^2)A_{1}^2\kappa} {6} ]dt
\right]\right).
\end{equation}

Here, $\gamma(t)<0$ and $P=(2-m^2)>0$. For, $m^2=1$, we get bell
shaped solitary wave solution as,
\begin{equation}
u=A_{1} dn\left( \sqrt {\frac{-A_{1}^2\kappa} {6}} \left[x-\int
[\alpha(t)- \frac {\gamma(t) A_{1}^2\kappa} {6} ]dt
\right]\right).
\end{equation}

In conclusion, MKdV equation with time varying coefficients has
solitary waves solutions, provided the temporal variations of the
coefficients are of the form given in the text. The temporal
variation of these parameters allow effective control of the
solitary wave profile. These  continuous waves and localized
solutions can be made to accelerate. The amplitude and widths are
not modulated by the distributed coefficients. The induction of
time dependent $u_{x}$ term allows us to control the motion of the
solitons more efficiently.

\end{document}